\documentclass[aps,physrev,reprint,amsmath, amssymb,groupedaddress]{revtex4-2}
\usepackage{bm}
\usepackage{graphicx}
\usepackage{makecell}
\usepackage{multirow} 
\usepackage{xfrac}
\graphicspath{{images/}}
\newcommand{\PP}{\mathrm{p}}
\newcommand{\BB}{\mathrm{b}}
\newcommand{\TT}{\mathrm{T}}
\newcommand{\RR}{\mathrm{R}}
\newcommand{\largetau}{\scalebox{1.44}{$\tau$}}
\begin{document}

%
%
%
\preprint{APS/123-QED}
\title{\textbf{Impact of a tunneling particle on the quantum state of a barrier: two-particle wave-packet model}
}
\author{Roman Michelko}
\author{Peter Bokes}%
 \email{Contact author: peter.bokes@stuba.sk}
\affiliation{%
Department of Physics,
Institute of Nuclear and Physical Engineering, \\
Faculty of Electrical Engineering and Information Technology,   
Slovak University of Technology in Bratislava, 841 04 Bratislava, Slovak Republic
}
\date{\today}
\begin{abstract}
We investigate the scattering of two distinguishable particles with unequal masses and a mutual short-range interaction with the aim of quantifying the impact of a tunneling ``projectile'' particle on the quantum mechanical state of the ``barrier'' particle. 
We find that the states of the barrier particle after the tunneling or reflection of the projectile are displaced by a finite distance that is given by the derivative of the phase of the transmission or reflection amplitudes multiplied by factors dependent on particles' masses, respectively. We demonstrate these results on a numerical example with a resonant interaction between a projectile and barrier. Our work demonstrates physical implication of the concept of phase time delay in the form of finite displacements of particles that are, in principle, experimentally measurable.
\end{abstract}

%
%
%
\maketitle
%
\section{Introduction} \label{sec:intro}

Almost a hundred years ago, Darwin explored the use of Gaussian wave packets to describe the motion of a single free quantum particle~\cite{Darwin1927}
and introduced two-particle Gaussian-like wave packets to study the scattering of two quantum particles~\cite{Darwin1929}.
His motivation for the latter work was to demonstrate the emergence of particle-like behavior from wave mechanics, particularly in the case of
a heavier projectile particle colliding with a lighter target particle initially at rest.

Although most subsequent developments in scattering theory relied on the stationary scattering eigenstates, 
influential contributions employing wave packets also emerged. MacColl used wave packets to analyze quantum tunneling through a potential barrier~\cite{MacColl1932},
concluding that the wave function appears at the far end of the barrier as soon as it gains an appreciable amplitude at the entrance.
Wigner~\cite{Wigner1955}, building on his previous work with Eisenbud, interpreted the phase shift in the scattering wave in terms of time delay,
employing the stationary phase approximation for a general wave packet. Later, advances in the fabrication of ultra-thin metal-oxide-metal interfaces motivated
Hartman~\cite{Hartman1962} to revisit MacColl’s work. Using an analysis based on the scattering of a Gaussian wave packet on a square potential barrier and Wigner's phase time delay, he demonstrated that the time for a particle to tunnel through a thick barrier is shorter than the time a free packet would take
to traverse the same distance. In the limiting case, this time becomes independent of the barrier width.

The calculation of tunneling time delays using wave packets was later questioned and critically reviewed by many researchers, perhaps most notably
by Landauer and Martin~\cite{Landauer1994}. Their objections to wave-packet-based approaches and, by extension, to phase shift-derived delay times stemmed
from the apparent arbitrariness of the wave-packet shape and its influence on the results, as well as the absence of a direct experimental method
for measuring time delay. Instead, they advocated for alternative characterizations of tunneling time, such as Larmor time, which employs
a non-zero magnetic field within the tunneling region~\cite{Buttiker1983, Suzuki2023}, or traversal time~\cite{Buttiker1982}, where a weakly oscillating
barrier height with variable frequency of oscillation being a control parameter.
These approaches rely on the interaction of the tunneling particle with another dynamical system—either quantum or classical—commonly referred to as a "clock."

The idea that tunneling time should be determined through interaction with a clock was formalized within the framework of weak measurements
by Steinberg~\cite{Steinberg1995a,Steinberg1995b}. The relationship between the duration of an interaction process and quantum measurement
can be also traced back to the time-energy uncertainty relation~\cite{Aharonov1961} and its various interpretations through different forms of time operators~\cite{Maquet2014, Dodonov2015}.
These approaches lead to probability distributions for time delays rather than a single time scale~\cite{Kijowski1974,Galapon2018,Bokes2011}.

As emphasized by Landauer, interaction-based definitions of tunneling time have the advantage of suggesting experimental procedures for
their determination, as demonstrated by Gueret et al.\cite{Gueret1988} or recently by Ramos et al.\cite{Ramos2020}.
However, advances in quantum tomography have enabled also direct measurement of time delays for Gaussian wave packets~\cite{Kim2024},
potentially allowing time-of-flight methods to be used in semiconductor nanostructures~\cite{Park2023}.

The concept of phase-time delay gained prominence with the advent of attosecond photoemission spectroscopy in atoms~\cite{Schultze2010, Sainadh2020}.
In the case of photoemission, rather than studying an incoming wave packet, researchers analyze tunneling from a quasi-bound state or a resonance.
In this context, both ``clock'' time~\cite{Jia2022} and wave-packet bandwidth-averaged phase-time delays~\cite{Yusofsani2020} have been theoretically discussed.
Consequently, the debate over which definition of time delay is most appropriate in different circumstances remains ongoing~\cite{Rivlin2021, Ianconescu2021, Sokolovski2021}.

In our work, our starting point is similar to Darwin's: wave-packet formulation of scattering of two quantum particles in the laboratory frame of reference,
where a lighter projectile particle has a non-zero average velocity, while the target heavier particle has zero average velocity. Our aim is to understand how the projectile
influences the state of the target in the case of tunneling.
The interaction potential between the two particles represents the potential barrier for the projectile. The larger but finite mass of the target particle characterizes the dynamical degree of freedom of the barrier. For this reason, we refer to the target particle as the ``barrier'' particle.
In relation to the measurement of tunneling time using a ``clock'', this degree of freedom of the barrier serves as the measurement apparatus.
Physically, we expect that the time delay present in the phase shift will determine the behavior of the barrier particle after backscattering or tunneling.
In terms of the energy eigenstates, the problem is analytically solvable in the center-of-mass coordinate system as a one-particle scattering process, but the discussion of the two-particle finite wave-packet dynamics requires numerical computations. 
Our most significant result is that the phases of transmission and reflection coefficients lead to distinct spatial translations of the target particle after scattering, with qualitatively different outcomes for the two cases, respectively. 

The physical and mathematical model is introduced in Section II. Section III describes the construction of the time-dependent solution, and in Section IV the
analytical results that follow from the stationary phase approximation for long times are obtained. In Section V we introduce a model interaction that is used in our numerical implementation discussion is in Section VI.
Finally, in Section VII, we discuss the numerical results and compare them to the predictions of the stationary phase method. 

\section{\label{sec:iniWF}Initial state in laboratory and center of mass reference frames}
%
%
We consider two non-identical particles moving in one dimension along the $x$ axis. 
The lighter particle with mass $m_\PP$ will be referred to as the ``projectile," and the heavier particle with mass $m_\BB$ will be called the ``barrier.'' 
Accordingly, all physical quantities indexed by the lowercase letter ``p'' are related to the projectile, and those indexed by the lowercase letter ``b'' are related to the barrier. We choose the initial state of the two-particle system in the form of a moving Gaussian state, 
\begin{eqnarray}
\Psi_0(x_\PP, x_\BB) = &&\frac{1}{\sqrt{2\pi\sigma_{x,\PP} \sigma_{x,\BB}}} \exp \Bigg( 
 -\frac{(x_\PP - x_{\PP 0})^2}{4\sigma_{x,\PP}^2} \nonumber \\
  &&- \frac{(x_\BB - x_{\BB 0})^2}{4\sigma_{x,\BB}^2}
 + ik_{\PP 0} x_\PP + ik_{\BB 0} x_\BB \Bigg), \label{eq:initial_condition_LF}
\end{eqnarray}
where the particles have a normal distribution with the average positions $x_{\text{p}0}$ and $x_{\text{b}0}$,  the corresponding uncertainties being $\sigma_{x,\text{p}}$ and $\sigma_{x,\text{b}}$ respectively. The initial average velocities of the particles are $v_{\PP 0} = \hbar k_{\text{p}0}/m_\PP$ and $v_{\BB0} = \hbar k_{\text{b}0}/m_\BB$.
%
%

For a faithful representation of a collision the parameters of the initial state (\ref{eq:initial_condition_LF}) must fulfill two requirements. 
Firstly, prior to collision the projectile particle must be found almost certainly left of the barrier particle which demands:
\begin{equation} \label{eq:initial_pos_condition}
\sqrt{\sigma_{x, \PP}^2 + \sigma_{x, \BB}^2} \ll x_{\BB 0} - x_{\PP 0} . 
\end{equation}
The second necessary condition is that the velocity of the projectile is greater than that of the barrier $v_{\PP} > v_{\BB}$. For this to be satisfied with sufficiently large probability, the following inequality must be satisfied
\begin{equation} \label{eq:initial_vel_condition}
\sqrt{\sigma_{v, \PP}^2 + \sigma_{v, \BB}^2} \ll v_{\PP 0} - v_{\BB 0},
\end{equation}
where the uncertainities of the velocities follow from the form of the Gaussian initial state, $\sigma_{v, \PP}=\hbar/(2 m_{\PP} \sigma_{x, \PP})$
and $\sigma_{v, \BB}=\hbar/(2 m_{\BB} \sigma_{x, \BB})$.

In this work, we will consider the case where the projectile initially positioned to the left of the barrier $x_{\PP 0} < x_{\BB 0}$ with a positive
initial mean velocity $v_{\PP 0} > 0$ collides with the barrier having its initial mean velocity equal to zero $v_{\BB 0} = 0$. 
We fulfill the stated requirements by choosing a small position uncertainty compared to the initial mean distance between the particles and 
sufficiently large average initial velosity of the projectile. 
%

%
%
%
The two particles interact with each other via a short-range interaction prescribed by the potential $V(|x_\PP - x_\BB|)$ so that the Schrödinger equation
is separable in the center of mass (CoM) frame of reference.
With the introduction of two dimensionless mass ratios $\alpha = m_{\PP} / (m_{\PP} + m_{\BB})$ and $\beta = m_{\BB} / (m_{\PP} + m_{\BB})$, 
the transformation of particles' coordinates is
\begin{equation} \label{eq:CoM_transform}
\begin{pmatrix}
x_\PP\\ 
x_\BB
\end{pmatrix}
=
\begin{pmatrix}
1 & \beta\\ 
1 & -\alpha 
\end{pmatrix}
\begin{pmatrix}
X\\ 
x
\end{pmatrix}
\end{equation}
where $X$ refers to the position of CoM  
and $x$ is the relative position of the projectile with respect to the barrier. 
The initial-state wave function is expressible in terms of CoM frame coordinates accordingly, 
\begin{eqnarray} \nonumber
  \Psi_0(X, x)= &&\frac{1}{\sqrt{2\pi\sigma_{x,\PP} \sigma_{x,\BB}}}
  \exp\left(-\frac{1}{4}
    \begin{pmatrix}
   	  X - X_0, & x - x_0
    \end{pmatrix}
     \cdot
    \bm{\sigma}_{X,x}^{-1}  \right. \\
    &&\left. \cdot
    \begin{pmatrix}
     X - X_0 \\
     x - x_0
    \end{pmatrix}
    +iK_0X + ik_0x
  \right), \label{eq:inital_wavefunction_CoM}
\end{eqnarray}
where $\bm{\sigma}_{X,x}$ is a non-diagonal covariance matrix
\begin{equation} \label{eq:Xx_cov_matrix}
\bm{\sigma}_{X,x} \equiv 
\begin{pmatrix}
 \alpha^2 \sigma_{x,\PP}^2 + \beta^2 \sigma_{x, \BB}^2 & \alpha \sigma_{x, \PP}^2 - \beta \sigma_{x, \BB}^2 \\
 \alpha \sigma_{x, \PP}^2 - \beta \sigma_{x, \BB}^2 & \sigma_{x, \PP}^2 + \sigma_{x, \BB}^2
\end{pmatrix}.
\end{equation}
$X_0$ and $x_0$ are the initial average coordinate of CoM and the initial relative coordinate given by the inverse transform of (\ref{eq:CoM_transform})
\begin{equation} \label{eq:CoM_transform_0}
\begin{pmatrix}
X_0\\ 
x_0
\end{pmatrix}
=
\begin{pmatrix}
\alpha & \beta\\ 
1 & -1 
\end{pmatrix}
\begin{pmatrix}
x_{\PP0}\\ 
x_{\BB0}
\end{pmatrix}
\end{equation}
and finally the average transformed wavenumbers are given by the transpose of (\ref{eq:CoM_transform})
\begin{equation} \label{eq:CoM_transform_K}
\begin{pmatrix}
K_0\\ 
k_0
\end{pmatrix}
=
\begin{pmatrix}
1 & 1\\ 
\beta & -\alpha 
\end{pmatrix}
\begin{pmatrix}
k_{\PP0}\\ 
k_{\BB0}
\end{pmatrix}
\end{equation}

%
%
Since the covariance matrix in CoM is not diagonal, the transformation into CoM induces correlations between the
positions of the particles in this frame of reference. 
%

%
%
\section{Time evolution of two-particle wave packet} 
In CoM frame of reference the solution for the two-particle Schrödinger equation
\begin{equation} \label{eq:complete_SchE}
i\hbar\frac{\partial \Psi}{\partial \tau} = -\frac{\hbar^2}{2M} \frac{\partial^2 \Psi}{\partial X^2} - \frac{\hbar^2}{2\mu} \frac{\partial^2 \Psi}{\partial x^2} + V(x)\Psi,
\end{equation}
where $M=m_\PP + m_\BB$ is the total mass of the system and $\mu=m_\PP m_\BB / M$ the reduced mass,
can be constructed using the standard separation technique.
%
\begin{equation} \label{eq:temporal_solution}
 \Psi(X, x, \tau) = \Phi(X)\psi(x)e^{-i E \tau / \hbar},
\end{equation}
where the motion of CoM is given by the plane wave $\Phi_K(X)=e^{iKX}$, $K \in \mathbb{R}$ and the state of the relative motion is given by the right- or left-going scattering 
states $\psi_{\gamma,k}(x)$, $k \in (0,\infty)$ and $\gamma \in \{\mathrm{R}, \mathrm{L} \}$. 
Outside the interval $x \in (-d/2, d/2)$, where interaction $V(x) = 0$, the scattering states can be expressed as:
\begin{eqnarray} 
&&\psi_{\text{R}, k}(x) =
    \begin{cases} 
        e^{ikx} + r_{\mathrm{R},k} e^{-ikx}, & x < -d/2 \\[.7em] 
         t_{\mathrm{R},k} e^{ikx}, & x > d/2
    \end{cases}, \label{eq:partial_psi_Rk_state} \\
&&\psi_{\text{L}, k}(x) =
    \begin{cases} 
        t_{\mathrm{L},k} e^{-ikx}, & x < -d/2 \\[.7em] 
        e^{-ikx} + r_{\mathrm{L},k} e^{ikx}, & x > d/2
    \end{cases}, \label{eq:partial_psi_k_state}
\end{eqnarray}
where $t_{\gamma, k}, r_{\gamma, k}$ are the amplitudes of reflection and transmission, dependent on the specific form of the potential~$V(x)$.
The total energy of the state (\ref{eq:temporal_solution}) is given by two contributions $E = E_K + E_k$  where
$E_K = \hbar^2K^2/(2M)$ and $E_k=\hbar^2k^2/(2\mu)$.

We can construct the wave packet corresponding to the system of two particles that comply with the initial state (\ref{eq:inital_wavefunction_CoM}) within the complete basis of stationary eigenstates $\psi_{\gamma,k}(x)$ and $\Phi_K(x)$. This allows us to seek a time-dependent solution to the problem in terms of integral over all possible stationary states.
\begin{eqnarray} \nonumber
 \Psi(X, x, \tau) =&& \frac{1}{2\pi}\sum_\gamma \iint \mathrm{d}K\mathrm{d}k \, c_\gamma(K, k) \Phi_K(X) \\
 &&\times \psi_{\gamma, k}(x)e^{-iE\tau/\hbar} \label{eq:wavefunction}
\end{eqnarray}
With knowledge of the initial wave function, the amplitude coefficient function $c_{\gamma}(K, k)$ can be found by evaluating the integral
\begin{equation} \label{eq:coeff_func}
 c_\gamma(K, k) = \frac{1}{2\pi}\iint \mathrm{d}X \mathrm{d}x \, \Psi_0(X, x)\Phi_K^*(X)\psi_{\gamma, k}^*(x).
\end{equation}
Assuming the solutions $\psi_{\gamma, k}(x)$ in the form given in (\ref{eq:partial_psi_k_state}), we see that due to the branching the integration along the whole $x$ axis  splits into two intervals:
\begin{eqnarray} 
\int \mathrm{d}x \ldots = \int_{-\infty}^{-d/2} \mathrm{d}x \ldots + \int_{-d/2}^{\infty} \mathrm{d}x \ldots  \label{eq:c_integral_x}
\end{eqnarray}
%
%
As we discussed in Sec.~\ref{sec:iniWF}, the initial state $\Psi_0(X, x)$  
is strongly localized at $x = x_0 = x_{\PP0}-x_{\BB0}$ to satisfy the requirement of no overlap (\ref{eq:initial_pos_condition}). This implies that it will be exponentially small for $x \geq -d/2$ and the only significant contribution to the amplitude function $c_{\gamma}(K, k)$ is due to the first integral in (\ref{eq:c_integral_x}). 

From the expression for $\psi_\mathrm{L,k}(x)$ and $\psi_\mathrm{R,k}(x)$, we can see that the final form of the amplitude coefficient functions for the left-propagating and right-propagating states consists of two parts:
\begin{gather}
c_{\gamma}(K, k) = 
\begin{cases}
	t_{\mathrm{L},k}^*C(K, -k), & \quad \gamma = \mathrm{L}, \\
	C(K, k) + r_{\mathrm{R},k}^*C(K, -k), & \quad \gamma = \mathrm{R}. 
\end{cases}
\end{gather}
where we have introduced the function $C(K, k)$
\begin{equation}
C(K, k) = \frac{1}{2\pi} \iint\limits_{x < -d/2} \mathrm{d}x \mathrm{d}X \, \Psi_0(X,x)e^{-iKX}e^{-ikx}.
\end{equation}
defined for $k \in \mathbb{R}$.
$C(K, k)$ can be expressed in a closed but rather elaborate analytical form. However, its major contribution comes from 
a simple Gaussian parts and phase factors:
\begin{eqnarray}
C(K, k) \approx \,
&& \exp\Bigg( 
    -\frac{(k_\PP - k_{\PP 0})^2}{4\sigma_{k, \PP}^2}
    -\frac{(k_\BB - k_{\BB 0})^2}{4\sigma_{k, \BB}^2} \Bigg. \nonumber \\
&& \Bigg. -\, i x_{\PP 0} k_\PP - \, i x_{\BB 0}k_\BB \Bigg), \label{eq:CkPkB}
\end{eqnarray}
where
\begin{equation} \label{eq:Kk_transform}
\begin{pmatrix}
k_\PP\\ 
k_\BB
\end{pmatrix}
=
\begin{pmatrix}
\alpha & 1\\ 
\beta & -1 
\end{pmatrix}
\begin{pmatrix}
K\\ 
k
\end{pmatrix}
\end{equation}
The approximate form (\ref{eq:CkPkB}) is actually the exact 2D Fourier transform of the initial state $\Psi_0(k_\PP,k_\BB) = \mathcal{F}[\Psi_0(x_\PP,x_\BB)]$ if performed within the laboratory frame of reference, as one would expect. 
In the explicit form of Eq.~(\ref{eq:CkPkB}) we can identify the wavenumber covariance matrix $\bm{\sigma}_{K,k}$, in close
analogy with the derivation of (\ref{eq:inital_wavefunction_CoM}) from (\ref{eq:initial_condition_LF}).

We can see that $C(K, k)$ quickly tends to zero for $|K - K_0| \gg \sigma_K$ and $|k - k_0| \gg \sigma_k$, where $\sigma_K$ and $\sigma_k$ are square roots of the diagonal elements of the initial wave-number covariance matrix $\bm{\sigma}_{K,k}$. 
From the requirement (\ref{eq:initial_vel_condition}) follows that $k_0 \ll \sigma_{k}$, which in turn implies that the function $C(K, -k)$ is negligible for all $k>0$.
Hence  $c_\text{L}(K, k) \approx 0$ and 
$c_\gamma(K, k)=C(K,k) \delta_{\mathrm{R},\gamma}$ is independent of the choice of interaction potential.
This confirms the expectation that for a well-posed initial setup, the left-propagating states play no significant role in the system dynamics. 
For this reason, from this point on, we only assume the right-going states omitting the index $\gamma=\text{R}$. 
\section{Asymptotic results for particles' positions} \label{sec:StatPhase}

Our primary objective is to study the change in the state of the barrier as a consequence of a tunneling projectile. 
At large times, $\tau \gg |x_0|/v_0$, the two-particle wave packet will be non-zero only for $|x| \gg d/2$. For $x \ge d/2$, using Eqs.~(\ref{eq:wavefunction}), (\ref{eq:partial_psi_Rk_state}) we obtain the transmitted wave
\begin{equation} \label{eq:tunneling_wavefunction}
\Psi(X,x\geq d/2,\tau) = \frac{1}{2\pi}\iint \mathrm{d}K\mathrm{d}k \, | C(K, k) t_k| e^{i\phi(k,K,\tau)}
\end{equation}
where the phase for the transmitted wave is 
\begin{eqnarray} \nonumber
 \phi(K, k) = &&K(X-X_0) + k(x-x_0) - \frac{\hbar K^2}{2M}\tau - \frac{\hbar k^2}{2\mu}\tau \\
 &&+ \varphi_t(k), \label{eq:phase_function}
\end{eqnarray}
and $\varphi_t(k)$ is the phase of the transmission amplitude $t_k = |t_k| e^{i\varphi_t(k)}$.

We can now employ the stationary phase method~\cite{Wigner1955, Bohm1951} for asymptotic behaviour for the two-particle wave packet. The dominant contributions to the integral come from $K_\text{max}$ and $k_\text{max}$ for which $|C(K,k) t_k|$ attains maximum~\cite{Bernardini2007}. 
As time increases, the magnitude of the phase $\phi(K,k,\tau)$ increases , resulting in rapid oscillations of the exponential factor. These oscillations diminish the wave-packet amplitude for all $x, X$ except when they are close to such $x(\tau), X(\tau)$ for which the phase is stationary at $K_\text{max}$ and $k_\text{max}$, 
i.e. 
\begin{eqnarray} \nonumber
 \frac{\partial \phi}{\partial K} \bigg|_{K_{\text{max}}} &=& X(\tau) - X_0 - \frac{\hbar K_{\text{max}}}{M} \tau  =0, \\ 
 \frac{\partial \phi}{\partial k} \bigg|_{k_{\text{max}}} &=& x(\tau) - x_0 - \frac{\hbar k_{\text{max}}}{\mu}\tau + \frac{\partial\varphi_t}{\partial k}\bigg|_{k_{\text{max}}}= 0.
\end{eqnarray}
Using the transformation back to the laboratory frame (\ref{eq:CoM_transform}), (\ref{eq:CoM_transform_0}) we find the positions of the projectile and the barrier in the case that the projectile tunneled through the barrier:
\begin{eqnarray} \nonumber
 &x_{\PP}(\tau) =& x_{\PP 0}+\tau v_{\PP \infty} - \beta \frac{\partial \varphi_t}{\partial k} \bigg|_{k_{\text{max}}}, \\
 &x_{\BB}(\tau) =& x_{\BB 0}+\tau v_{\BB \infty} + \alpha \frac{\partial \varphi_t}{\partial k} \bigg|_{k_{\text{max}}}. \label{eq:sp_tunneling}
\end{eqnarray}
where $v_{\PP \infty}$ and $v_{\BB \infty}$ are similarly expressed using the transformation 
of the velocity of CoM and the relative velocity
\begin{equation} \label{eq:v_infty}
\begin{pmatrix}
v_{\PP \infty} \\ 
v_{\BB \infty} 
\end{pmatrix}
=
\begin{pmatrix}
1 & \beta\\ 
1 & -\alpha 
\end{pmatrix}
\begin{pmatrix}
\hbar K_{\text{max}}/M \\ 
\hbar k_{\text{max}}/\mu
\end{pmatrix}
\end{equation}

For a wave packet with small uncertainties in the initial velocities of the projectile and barrier, the function $|C(K,k)|$ will have a sharp maximum at $K_0,k_0$. From this it follows that $K_\mathrm{max} \approx K_0, k_\mathrm{max} \approx k_0$ and 
hence using Eq.~(\ref{eq:v_infty}) and Eq.~(\ref{eq:CoM_transform_K}) we find
$v_{\PP \infty} \approx v_{\PP 0}$ and $v_{\BB \infty} \approx v_{\BB 0}$, as if there were
no collision at all. However, this is not precisely true when we inspect the kinematics Eq.~(\ref{eq:sp_tunneling}) as both projectile and the barrier acquire an additional shift in their positions,
\begin{equation} \label{eq:barrier_displacement}
\Delta x_{\BB, \TT} = \alpha \frac{\partial \varphi_t}{\partial k} \bigg|_{k_{\text{max}}}, \quad 
\Delta x_{\PP, \TT} = - \beta \frac{\partial \varphi_t}{\partial k} \bigg|_{k_{\text{max}}}
\end{equation}
where we have added the index ``T'' to indicate that these shifts correspond to the case when we observe tunneling. 

These results are particularly striking when the barrier particle is initially at rest, $v_{\BB 0} = 0$. According to Eq.~\eqref{eq:sp_tunneling} and the discussion above it remains at rest also after the tunneling event, but its final position is displaced by $\Delta  x_{\BB,\TT}$.  In general, for an incident wave packet with finite widths (or finite uncertainities in initial velocities)
the velocity of the barrier after collision will not be precisely zero since $k_\mathrm{max} \neq k_{0}$ and $K_\mathrm{max} \neq K_{0}$ due to the dependence of the absolute value of the tunneling amplitude on $k$. 
We will demonstrate these results on a numerical example with a resonant interaction between the projectile and barrier in later sections.

%
%

The stationary phase approximation with analogous sequence of steps can be applied to the reflected part of the wave packet with the overall phase
\begin{eqnarray} \nonumber
 \phi_\RR(K, k) = &&K(X-X_0) + k(-x-x_0) - \frac{\hbar K^2}{2M}\tau - \frac{\hbar k^2}{2\mu}\tau \\
 &&+ \varphi_r(k). \label{eq:phase_function_R}
\end{eqnarray}
where $\varphi_r(k)$ is now the phase of the reflection amplitude $r_k = |r_k| e^{i\varphi_r(k)}$.
This results in the asymptotic kinematics of the particle and the barrier 
\begin{eqnarray} \nonumber
    &x_{\PP, \RR}(\tau) =& x_{\PP, \mathrm{v}} + \tau v_{\PP\infty, \RR} + \beta \frac{\partial \varphi_r}{\partial k} \bigg|_{k_{\text{max}, \RR}}, \\
    &x_{\BB, \RR}(\tau) =& x_{\BB, \mathrm{v}} + \tau v_{\BB \infty, \RR} -\alpha \frac{\partial \varphi_r}{\partial k} \bigg|_{k_{\text{max}, \RR}}, \label{eq:sp_scattering}
\end{eqnarray} 
where $x_{\PP, \mathrm{v}}$, $x_{\BB, \mathrm{v}}$ are the classical ``virtual initial positions'' describing the kinematics after the collision of two point-like particles:
%
\begin{equation} \label{eq:virtual_pos}
\begin{pmatrix}
x_{\PP,\text{v}}\\ 
x_{\BB,\text{v}}
\end{pmatrix}
=
\begin{pmatrix}
1 & \beta\\ 
1 & -\alpha 
\end{pmatrix}
\begin{pmatrix}
1 & 0 \\ 
0 & -1 
\end{pmatrix}
\begin{pmatrix}
\alpha & \beta\\ 
1 & -1 
\end{pmatrix}
\begin{pmatrix}
x_{\PP0}\\ 
x_{\BB0}
\end{pmatrix}
\end{equation}
The asymptotic velocities after the reflection, $v_{\PP \infty, \RR}$, $v_{\BB \infty, \RR}$ depend on the maximum of the reflected-wave amplitude $|C(K, k) r_k|$ located at $K_\mathrm{max}, k'_{\text{max}}$,
\begin{equation} \label{eq:v_R_infty}
\begin{pmatrix}
v_{\PP \infty,\RR} \\ 
v_{\BB \infty, \RR} 
\end{pmatrix}
=
\begin{pmatrix}
1 & \beta\\ 
1 & -\alpha 
\end{pmatrix}
\begin{pmatrix}
\hbar K_{\text{max}}/M \\ 
- \hbar k'_{\text{max}}/\mu
\end{pmatrix}
\end{equation}
Similarly to the discussion in the case of tunneling, for narrow a wave packet $k'_\mathrm{max} \approx k_{0}$ and $K_\mathrm{max} \approx K_{0}$ and the asymptotic velocities are related to the initial ones using the composite transformation identical to the one for virtual positions (\ref{eq:virtual_pos}), 
$v_{\PP \infty, \RR} = (\alpha - \beta) v_{\PP 0} + 2\beta v_{\BB 0}$ and $v_{\BB \infty, \RR} = 2\alpha v_{\BB 0} - (\alpha - \beta) v_{\BB 0} $. This is in agreement with the result for an elastic collision in one dimension in classical mechanics. The details of the two-particle backscattering slowdown and acceleration are contained in the displacements
\begin{equation} \label{eq:barrier_displacement_R}
\Delta x_{\BB, \RR} = - \alpha \frac{\partial \varphi_r}{\partial k} \bigg|_{k'_{\text{max}}}, \quad 
\Delta x_{\PP, \RR} =  \beta \frac{\partial \varphi_r}{\partial k} \bigg|_{k'_{\text{max}}}.
\end{equation}

We note the difference in the barrier displacement in the tunneling (\ref{eq:barrier_displacement}) and reflection (\ref{eq:barrier_displacement_R}) so that these can be discerned from each other and also from the admissible case that the projectile would not reach the barrier particle at all, when the displacement would be simply zero.

%
%
%
\section{Specification of a two-particle model with resonant interaction}

In the previous section, we have obtained asymptotic results for the displacements of the particles after their collision, which differ for the projectile experiencing backscattering or tunneling. These are valid for a narrow two-particle wave packet and for asymptotic times. In this section, we explore to what extent these remain valid also for fairly localized wave functions and finite-time scales. 

Examining the expressions (\ref{eq:barrier_displacement}), we expect that barrier displacement is most significant, when the phase of the transmission coefficient vary substantially over the range of momenta in the wave packet. We can easily achieve this by choosing the resonant interaction potential with two symmetrically positioned delta function barriers~\cite{Bokes2011}
\begin{equation} \label{eq:interaction_potential}
 V(x) = V_0 \left(\delta(x+d/2) + \delta(x-d/2)\right)
\end{equation}
where $d$ sets the short range of particles' interaction or, from a perspective of the relative motion, 
the width of the interaction potential and $V_0$ adjusts the interaction strength.
The coefficients of reflection and transmission corresponding to the resonant interaction potential \eqref{eq:interaction_potential} are given by the formlae
\begin{eqnarray} \nonumber
    &r_k=&\frac{2V_0\mu}{\hbar^2}\frac{\cos(kd)+\frac{V_0\mu}{\hbar^2}\sin(kd)}{\left(i-\frac{2V_0\mu}{\hbar^2} \right)e^{-ikd}-2\left( \frac{V_0\mu}{\hbar^2}\right)^2\sin(kd)}, \\
    &t_k=&\frac{1}{\left(1+i\frac{2V_0\mu}{\hbar^2} \right)e^{-ikd}+2i\left(\frac{V_0\mu}{\hbar^2}\right)^2\sin(kd)}, \label{eq:coefficients_rk_tk}
\end{eqnarray}
from which we also obtained an analytical expression for the relative wavenumber derivative of the transmission coefficient phase needed for the calculation of the displacements Eqs.~(\ref{eq:barrier_displacement}) and (\ref{eq:barrier_displacement_R}). For a symmetric potential, $V(x) = V(-x)$, the unitarity of the S-matrix implies that the phase of the reflection and the phase of the transmission coefficient may differ only by a constant, so that these two derivatives are equal for our form of the interaction.  

%
%
%
%
%
\begin{table}[t]
\caption{\label{tab:free_parameters}%
Independent free parameters $\alpha$, $\eta$, and $d$ for six selected cases, along with calculated transmission probabilities and wavenumber uncertainties. Cases are grouped based on the significance of point $T(k_0)$.
}
\begingroup
\small
\renewcommand{\arraystretch}{1.3}
\begin{ruledtabular}
\begin{tabular}{ccccccc}
\textrm{Case} &
\textrm{$\alpha$} &
\textrm{$\eta$} &
\textrm{$T(\%)$} &
\makecell{$d$\\$(\times 10^{-2})$} &
\makecell{$\sigma_k/k_0$\\$(\times 10^{-2})$} &
\textrm{Point $T(k_0)$} \rule[-3ex]{0pt}{0pt} \\
\colrule
1 & $\sfrac{1}{101}$ & \multirow{2}{*}{24} & 50.20 & \multirow{2}{*}{6.552} & 4.95 & \multirow{2}{*}{First minimum} \\
2 & $\sfrac{1}{3}$   &                     & 49.83 &                        & 3.73 &                             \\
\hline
3 & $\sfrac{1}{101}$ & \multirow{2}{*}{24} & 98.43 & \multirow{2}{*}{4.037} & 4.95 & \multirow{2}{*}{First maximum} \\
4 & $\sfrac{1}{3}$   &                     & 99.10 &                        & 3.73 &                              \\
\hline
5 & \multirow{2}{*}[-1.8mm]{$\sfrac{1}{3}$} & \multirow{2}{*}[-1.8mm]{48} & 68.15 & 4.154 & \multirow{2}{*}[-1.8mm]{3.73} & \makecell{\rule[2.5ex]{0pt}{0pt} First rising\\point of inflection} \\
6 &                                &                      & 78.57 & 5.060 &                       & \makecell{First falling\\point of inflection} \\
\end{tabular}
\end{ruledtabular}
\endgroup
\end{table}
%
In our numerical calculations we use units that refer to the system's initial state. The set of three base units of length, time, and mass is defined. The unit of length $\ell$ is taken as the initial mean distance between two particles $|x_0|$. The unit of time $\largetau$ is selected as the time in which the particles would collide if the initial relative velocity was exactly $v_0 = v_{\PP 0} - v_{\BB 0}$. Finally, the unit of mass is taken as the reduced mass of the system $\mathrm{\mu}$. All other necessary units, such as velocity, momentum, or energy, can easily be derived from the three basic units. These units directly fix the mean initial relative velocity value to unity $v_0 = 1 \ \ell\largetau^{-1}$.
From now on all quantities and expressions are given in these dimensionless units, e.g.    
$x \to x/\ell$ or $k \to k/( \ell^{-1})$.
The Schrödinger equation after this scaling transformation takes form 
\begin{equation} \label{eq:nat_unit_SchE}
ik_0 \frac{\partial \Psi}{\partial t} = -\frac{\alpha \beta}{2} \frac{\partial^2 \Psi}{\partial X^2} - \frac{1}{2} \frac{\partial^2 \Psi}{\partial x^2} + V(x)\Psi , 
\end{equation}
where the dimensionless initial wavenumber 
$k_0 = \mu \ell^2/(\hbar\largetau)$
%
%
is inversly proportional to Planck's constant; hence the choice of its value directly affects the manifestation of all the observed quantum behavior.
The parameter $\eta = V_0 \mu \ell / \hbar^2$ sets the scale of interaction strength in our dimensionless unit system.


%

Choosing suitable units does not remove all the freedom of choice in the model. 
The initial particle uncertainties are set equally to the value $\sigma_{x, \PP} = \sigma_{x, \BB} = 0.2$, which is a value small enough to guarantee the condition (\ref{eq:initial_pos_condition}). We chose $k_0=50$ according to the requirements (\ref{eq:initial_vel_condition}). In table \ref{tab:free_parameters}, we present all other parameters that uniquely determine the six cases we considered. We explored two distinct situations based on the parameter $\alpha$; in four out of six scenarios, the value $\alpha=\sfrac{1}{3}$ signifies the ratio of mass 1:2 between the projectile and the barrier. In the remaining two cases, 2 and 4, the mass ratio of 1:100 with $\alpha = \sfrac{1}{101}$ represents a significantly lighter projectile. Two different values of the strength of interaction $\eta$ were used for the presented results. 

%
\begin{figure}[t]
    \centering
    \includegraphics[width=\columnwidth]{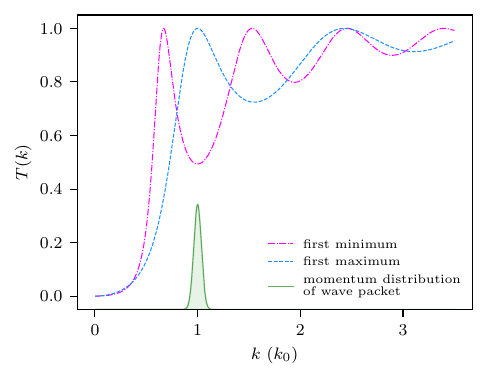}
    \caption{Functions $T(k)$ with $k_0$ located in its first minimum or maximum (Case 2 and Case 4).}
    \label{img:plot_T_k_1}
\end{figure}
%

\begin{figure}[b]
    \centering
    \includegraphics[width=\columnwidth]{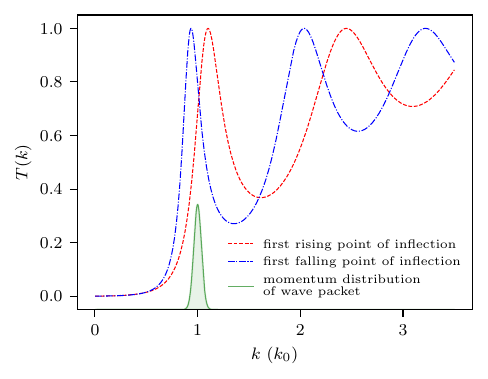}
    \caption{Functions $T(k)$ with $k_0$ located in the first rising and falling point of inflection (Case 5 and Case 6).}
    \label{img:plot_T_k_2}
\end{figure}

The last parameter $d$ is deliberately chosen so that $k_0$ is either at the extreme or the inflection point of the function $T(k)=|t_k|^2$. The function $T(k)$ for different values of $d$ is shown in figure \ref{img:plot_T_k_1}. For the first two cases,  $k_0$ is the first minimum of $T(k)$; in cases 3 and 4, it is the first maximum. With $k_0$ at minimum, the transmission probabilities are approximately 50 \%, while with $k_0$ at maximum, they are approximately 98-99 \%, while keeping the same barrier strength $\eta$. In the last two cases, the function $T(k)$ is depicted in the figure \ref{img:plot_T_k_2}. Here, the choice of $d$ places the points $k_0$ on the first rising and falling inflection point of $T(k)$.
%

\section{Numerical calculation of the wave-packet evolution}
Time development of the system can be obtained directly by numerical evaluation of the wave function (\ref{eq:wavefunction}) in CoM frame of reference. For the task, we have developed Python code utilizing 2-dimensional fast Fourier transform (FFT) to calculate the wave function on the discretized square grid of $(X, x)$ points in an arbitrary instance of time in the range $0 \leq \tau < \tau_\text{max}$. $\tau_\text{max}$ represents the maximum time threshold at which aliasing does not occur. 
%
%
\begin{figure}[!b]
    \centering
    \includegraphics[width=\columnwidth]{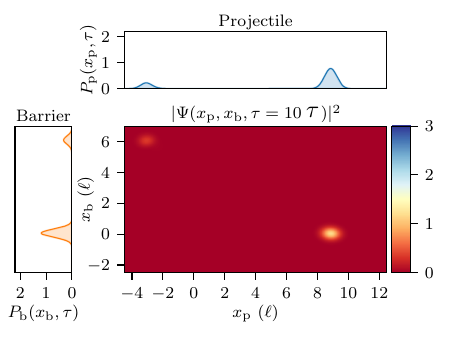}
    \caption{Center panel - Map of probability density function at time $\tau=10$, with transmitted portion of wave packet on the right and reflected portion in the top left. Top panel - marginal probability density for projectile. Left panel - marginal probability density for barrier.}
    \label{fig:wavepacket}
\end{figure}
%
%

More specifically, the discretized square grid represented a region of area $L \times L$ where $L=100$ with $N = 2^{12}$ discrete points in both $X$ and $x$ directions. With this choice, spatial points $(X, x)$ in our grid are sampled with the resolution $\Delta_{X,x} = L/N$, while the points in the $K, k$ space are sampled with the spatial frequency $\Delta_{K, k} = 2\pi / L$. The chosen initial particles' uncertainties are large enough compared to the discretization step $|x_0| \gg \sigma_x > \sigma_X \gg \Delta_{X,x}$. In combination, this choice of $N, L$ and $k_0$ results in a maximum time threshold of $\tau_\text{max} \approx 19.4$. All calculations were performed within the time interval $8 \leq \tau \leq 12$ using a time step of $\Delta \tau = 0.1$.

For the CoM wave function $\Psi(x, X, \tau)$ calculated on the grid for the desired times, we obtain the probability density in a laboratory frame $P(x_\PP, x_\BB, \tau)$ from the probability density in CoM $|\Psi(x, X, \tau)|^2$ using the inverse transformation~\eqref{eq:CoM_transform_0} in a point-by-point fashion. 

\begin{figure}[!t]
    \centering
    \includegraphics[width=\columnwidth]{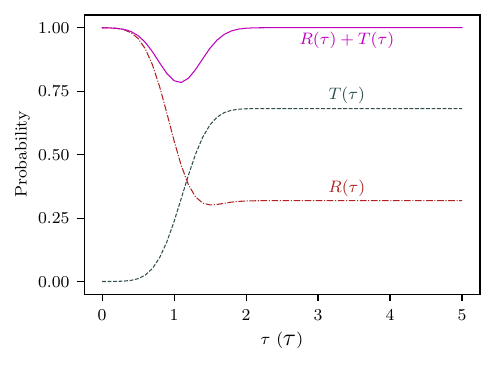}
    \caption{Evolution of conditional probabilities norms over time $0 \leq \tau \leq 5$ in numerical computation. White $T(\tau)$ has direct interpretation as the probability of tunneling, $R9(\tau)$ gives the probability of reflection only for $\tau \gg 1$.}
    \label{fig:norms}
\end{figure}

As an example, we show the numerically calculated probability density $P(x_\PP, x_\BB, \tau)$ for $\tau=10$ on the central panel of figure \ref{fig:wavepacket} and the corresponding animation for $\tau \in (0,12)$ is given as supplementary 
online material~
\footnote{In supplementary material we provide a short
video wavepacket-case6.mp4 with time evolution of the probability density
for a two-particle wave packet.}.
The specific parameters for this demonstration are those of the Case 6 in table~\ref{tab:free_parameters}. The reflected and transmitted portions of the initial wave packet are easily identifiable as they propagate away from the collision region $x_\PP \approx x_\BB \approx 0$. We can also track the motion of particles from marginal probability densities
\begin{eqnarray}
P_\PP(x_\PP,\tau) &=& \int \mathrm{d} x_\BB P(x_\PP,x_\BB,\tau), \\
P_\BB(x_\BB,\tau) &=& \int \mathrm{d} x_\PP P(x_\PP,x_\BB,\tau), 
\end{eqnarray}
visible in the left and top panels of the figure \ref{fig:wavepacket}.
The two-particle probability density clearly contains two 
entangled alternatives:
either the projectile particle has tunneled through the barrier and the barrier stayed essentially very close to the origin or the projectile was reflected back and the barrier gained momentum and is moving in the direction of the projectile's initial velocity. To separate them into
tunneling ($x_\PP > x_\BB$) and reflection ($x_\PP < x_\BB$ for $\tau \gg 1$) we introduce 
conditional probability densities:
\begin{eqnarray} \nonumber
&P_\TT(x_\PP, x_\BB, \tau) =& \theta(x_\PP - x_\BB) \frac{P(x_\PP, x_\BB, \tau)}{T(\tau)}, \\
&P_\RR(x_\PP, x_\BB, \tau) =& \theta(x_\BB - x_\PP) \frac{P(x_\PP, x_\BB, \tau)}{R(\tau)},\label{eq:projected_pdf}
\end{eqnarray} 
where $\theta(x)$ is a Heaviside step function and 
\begin{eqnarray} \nonumber
    &T(\tau) = &\iint\limits_{x_\PP > x_\BB} \mathrm{d}x_\PP \mathrm{d}x_\BB \, P(x_\PP,x_\BB,\tau), \\
    &R(\tau) = &\iint\limits_{x_\PP < x_\BB} \mathrm{d}x_\PP \mathrm{d}x_\BB \, P(x_\PP,x_\BB,\tau).  \label{eq:numerical_probability_of_tunneling}
\end{eqnarray}
$T(\tau)$ and $R(\tau)$ clearly have the meaning of the probability of tunneling (or transmission) and the probability of backscattering (or reflection) for large times respectively. Their evolution in time for Case 6 is shown in figure~\ref{fig:norms}. 
The noticeable drop in normalization around the time of collision is partly caused by the omission of the solution in a small interaction interval $|x_\PP - x_\BB| < d/2$, but more importantly by significant interference effects during the collision process. After the collision, the conditional probabilities settle at a constant value probability of reflection $R$ and transmission $T$.

The conditional probability densities finally allow for the averaging of particle's positions subject to tunneling or backscattering events, i.e. for $\tau \gg 1$:
\begin{eqnarray} \nonumber
   &\bar{x}_{\PP, \TT}(\tau) =& \iint \mathrm{d}x_\PP \mathrm{d}x_\BB \, x_\PP P_\mathrm{T}(x_\PP, x_\BB, \tau), \\ 
   &\bar{x}_{\BB, \RR}(\tau) =& \iint \mathrm{d}x_\PP \mathrm{d}x_\BB \, x_\BB P_\mathrm{R}(x_\PP, x_\BB, \tau). \label{eq:numerical_avg_pos} 
\end{eqnarray} 

\section{Comparison and discussion of the results}
\begin{figure}[t]
    \centering
    \includegraphics[width=\columnwidth]{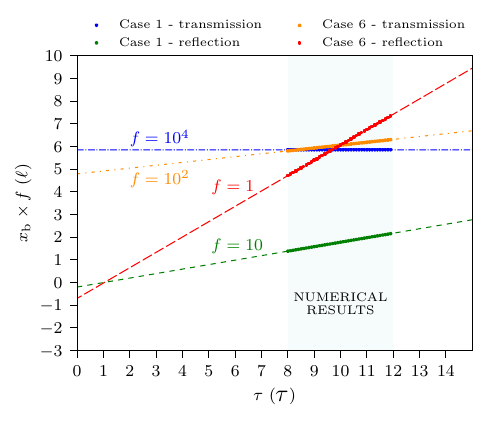}
    \caption{Numerical determined barrier trajectories extrapolated to initial time for cases 1 and 6. Note that different scale factors $f$ are needed to show the barrier displacements on the same axis.}
    \label{img:barrier_trajectories}
\end{figure}

We performed numerical calculations of two-particle probability densities for all six forenamed cases from table~\ref{tab:free_parameters}. Since we were primarily motivated by the question about the impact of the
tunneling particle on the barrier particle, we present quantitative results only for the latter.
We determined the barrier displacements $\Delta x_{\BB, \TT}$ and $\Delta x_{\BB, \RR}$ by linear extrapolation to $\tau=0$ from the numerically obtained average barrier positions in the chosen time frame from both the transmitted and reflected parts of the wave packet according to Eq.~\eqref{eq:numerical_avg_pos}. 
The extrapolated trajectories of the barrier in Cases 1 and 6 are demonstrated in figure \ref{img:barrier_trajectories}. These two cases illustrate two opposite extremes of barrier displacement: in 
Case 1, the barrier displacements are small, while in Case 6, they are the largest. The barrier velocity after the collision was determined from the slope of the average positions $x_{\BB, \TT}(\tau)$ and $x_{\BB, \RR}(\tau)$. 
%
%
%
%
The complete set of numerical results for the barrier particle, for all cases, along with the values predicted by the stationary phase method, is presented in tables \ref{tab:tunneling_results} and \ref{tab:scattering_results} assuming the transmission or reflection of the projectile particle, respectively. 
%
\begin{table*}[t]
\caption{\label{tab:tunneling_results}%
Comparison of $\Delta x_\BB$ and $v_\BB$ from stationary phase method and numerical computation in case of a tunneling projectile.}
\begin{ruledtabular}
\begin{tabular}{ccccccc}
\multirow{2}{*}{\textrm{Case}} & \multicolumn{2}{c}{$\Delta x_\BB$} & \multirow{2}{*}{$\delta_{\Delta x_\BB}$ (\%)} & \multicolumn{2}{c}{$v_\BB$} & \multirow{2}{*}{$\delta_{v_\BB}$ (\%)} \\
 & Stationary phase & Numerical & & Stationary phase & Numerical & \\ \hline
 \\[-1em]
1 & $5.74 \times 10^{-4}$ & $5.85 \times 10^{-4}$ & $1.82$ & $0$ & $1.52 \times 10^{-8}$ & -- \\
2 & $1.93 \times 10^{-2}$ & $1.95 \times 10^{-2}$ & $1.00$ & $0$ & $7.99 \times 10^{-6}$ & -- \\
3 & $7.74 \times 10^{-4}$ & $7.68 \times 10^{-4}$ & $0.71$ & $0$ & $-2.62 \times 10^{-8}$ & -- \\
4 & $2.61 \times 10^{-2}$ & $2.60 \times 10^{-2}$ & $0.39$ & $0$ & $-1.84 \times 10^{-5}$ & -- \\
5 & $4.16 \times 10^{-2}$ & $4.13 \times 10^{-2}$ & $0.70$ & $-1.59 \times 10^{-3}$ & $-1.74 \times 10^{-3}$ & $9.31$ \\
6 & $4.65 \times 10^{-2}$ & $4.79 \times 10^{-2}$ & $3.00$ & $1.32 \times 10^{-3}$ & $1.27 \times 10^{-3}$ & $3.92$ \\
\end{tabular}
\end{ruledtabular}
\end{table*}
%

%
\begin{table*}[t]
\caption{\label{tab:scattering_results}%
Comparison of $\Delta x_\BB$ and $v_\BB$ from stationary phase method and numerical computation in case of a backscattering projectile.}
\begin{ruledtabular}
\begin{tabular}{ccccccc}
\multirow{2}{*}{\textrm{Case}} & \multicolumn{2}{c}{$\Delta x_\BB$} & \multirow{2}{*}{$\delta_{\Delta x_\BB}$ (\%)} & \multicolumn{2}{c}{$v_\BB$} & \multirow{2}{*}{$\delta_{v_\BB}$ (\%)} \\
 & Stationary phase & Numerical & & Stationary phase & Numerical & \\ \hline
 \\[-1em]
1 & $-5.74 \times 10^{-4}$ & $-5.84 \times 10^{-4}$ & $1.68$ & $1.98 \times 10^{-2}$ & $1.98 \times 10^{-2}$ & $0.01$ \\
2 & $-1.93 \times 10^{-2}$ & $-1.95 \times 10^{-2}$ & $0.95$ & $6.67 \times 10^{-1}$ & $6.67 \times 10^{-1}$ & $0.00$ \\
3 & $-7.14 \times 10^{-4}$ & $-7.75 \times 10^{-4}$ & $8.44$ & $2.03 \times 10^{-2}$ & $1.95 \times 10^{-2}$ & $3.99$ \\
4 & $-2.45 \times 10^{-2}$ & $-2.61 \times 10^{-2}$ & $6.45$ & $6.79 \times 10^{-1}$ & $6.62 \times 10^{-1}$ & $2.47$ \\
5 & $-3.93 \times 10^{-2}$ & $-3.75 \times 10^{-2}$ & $4.53$ & $6.64 \times 10^{-1}$ & $6.58 \times 10^{-1}$ & $0.85$ \\
6 & $-4.11 \times 10^{-2}$ & $-3.86 \times 10^{-2}$ & $5.99$ & $6.71 \times 10^{-1}$ & $6.78 \times 10^{-1}$ & $1.03$ \\
\end{tabular}
\end{ruledtabular}
\end{table*}
%


First we consider displacements, for which we obtained good agreement between the stationary phase method and simulation in all the cases, the worst being the backscattering for Cases 3 and 4, where the relative differences are no more than 10\%. This discrepancy can be attributed to the fact that only a minuscule portion of around 1-1.5 \% of the initial wave packet was reflected in these cases. 
Between Cases 1 and 2 at the first minimum or Cases 3 and 4 at the first maximum, we observe that the relative differences improve slightly with heavier projectiles; this comes from the wave packet narrowing in momentum with an increasing value of $\alpha$.

Focusing on the velocities of the barrier when the projectile tunnels (Table~\ref{tab:tunneling_results}), we observe an interesting effect that results from the choice of point $k_\textrm{0}$ on the $T(k)$ curve. 
As we discussed in section~\ref{sec:StatPhase}, the asymptotic velocities of the tunneling particles are, in the case of a wave packet with narrow distribution of momentum, determined mainly by the behavior of function $T(k)$ in a neighborhood close to the point $k_0$. The asymmetry of $T(k)$ relative to the point $k_0$ will result in a shift of $k_\text{max}$ from $k_0$ in the transmitted wave packet, leading to an asymptotic velocity different from the initial one. 
Based on this argument, we expect that if the choice of parameters $d$ and $k_0$ results in an extremum of $T(k)$ at point $k_0$, as in Cases 1-4 then the numerically obtained asymptotic barrier velocities should be 
very close to zero in the case of tunneling. Although the values of barrier velocities in these cases are up to four orders of magnitude smaller than those of barrier displacements, they do not vanish entirely. 
This is due to the fact that the largest error in barrier velocity originating from the discretization of values $K$ and $k$ is on the order of $10^{-6}$ for $\alpha = \sfrac{1}{101}$ and $10^{-4}$ for $\alpha = \sfrac{1}{3}$.

Conversely, if we were to maximize the post-tunneling barrier velocity, we could choose the parameters $d$ and $\eta$ for $k_0$  to be at the steepest part of the curve $T(k)$, that is, the point of inflection as illustrated in Cases 5 and 6. Barrier velocities after projectile tunneling are in fact the most significant in these two cases. The sign of the barrier velocities depends on whether the function $T(k)$ is ascending or descending at $k_0$. We can once again observe this in Cases 5 and 6. In these cases, the asymptotic barrier velocity \( v_{\BB \infty} \) is negative for a rising point of inflection and positive for a falling point of inflection. 

As mentioned in the case of particle scattering off the barrier, the asymptotic velocities of particles can be shifted from the classical after-collision asymptotic velocities due to the maximum momentum shift in a reflected portion of the wave packet. The key point is that these shifts are significantly smaller than the recoil velocities of the barrier particle, making their contribution to the overall value relatively unimportant. This seems to be the reason for the exceptionally low differences between predicted and numerically obtained asymptotic barrier velocities after recoil.
The barrier's displacement is a telltale sign of a time scale of interaction between the two
particles. According to Wigner~\cite{Wigner1955}, the time delay $\Delta \tau_\phi = (\partial \phi_t / \partial k) v_0$ for an infinitely heavy barrier particle.  In our case of tunneling, if we assume that during the interaction the barrier was moving at an average speed $v_\BB$, the interaction time $\Delta \tau$ that results in displacement $\Delta x_\BB$ would be $\Delta \tau = \Delta x_\BB/v_\BB$. Similarly to $\Delta \tau_\phi$, $v_\BB$ is not a precisely defined quantity. Based on Newton's Third Law we expect $v_\BB \approx - (\alpha/\beta) \delta v_\PP$ i.e. $|v_\BB| \approx \alpha v_0$ for small to moderate $\alpha$. Hence, $\Delta \tau \approx (\partial \phi_t / \partial k) v_0 = \Delta \tau_\phi$. In more general situations, displacements~\eqref{eq:barrier_displacement}, \eqref{eq:barrier_displacement_R} characterize the finite duration of the interaction or tunneling event, can quantitatively differentiate between the two cases, and can be measured statistically by repeating a suitable scattering experiment. 

\section{Conclusions}

We investigated the scattering of two distinguishable particles with unequal masses and a mutual short-range interaction with the aim of quantifying the impact of a tunneling ``projectile'' particle on the quantum mechanical state of the ``barrier'' particle. Our investigation was motivated by a long-lasting discussion about physically relevant and measurable time scales in tunneling. The measurable impact of the tunneled projectile on the barrier could reveal the duration of the interaction and hence the tunneling process. We found that the state of the barrier particle after the tunneling or reflection of the projectile is displaced by a finite distance that is given by the derivative of the phase of the transmission or reflection amplitudes multiplied by factors dependent on particles' masses, respectively. Importantly, the two displacements are qualitatively and quantitatively different from each other, so the reflection or tunneling event can be discerned by conducting a measurement on the barrier particle only. 
Comparison of the analytical stationary phase results with numerical computation for a finite-width two-particle wave packet reassures us of the validity of the displacement formulae also for finite times. Our work demonstrates physical implication of the concept of phase time delay in the form of finite displacements of particles that are, in principle, experimentally measurable.

\begin{acknowledgments}
P.B. acknowledges the support of the Scientific Grant Agency VEGA under Project No. 2/0165/22 and
 the grant of Cultural and Educational Grant Agency (KEGA) of The Ministry of Education, Research, Development and Youth of the Slovak Republic No. 006STU-4/2025.
\end{acknowledgments}

%

\providecommand{\noopsort}[1]{}\providecommand{\singleletter}[1]{#1}%
\end{document}